\documentclass[onecolumn,11pt,a4]{IEEEtran}

\makeatletter
\def\ps@headings{%
\def\@oddhead{\mbox{}\scriptsize\rightmark \hfil \thepage}%
\def\@evenhead{\scriptsize\thepage \hfil \leftmark\mbox{}}%
\def\@oddfoot{}%
\def\@evenfoot{}}
\makeatother

\usepackage{color}
\usepackage{subfigure}
\usepackage{graphicx}
\usepackage{cite}
\usepackage{amssymb,amsfonts}
\usepackage{url}

\begin{document}
\title{Erasure Coding and Congestion Control for Interactive Real-Time Communication}


\author{
\IEEEauthorblockN{Pierre-Ugo Tournoux$^2$, Tuan Tran Thai$^1$, Emmanuel Lochin$^1$, J\'{e}r\^{o}me Lacan$^1$, Vincent Roca$^1$\\}
\IEEEauthorblockA{~\\$^1$Universit\'{e} de Toulouse, France ~~~ $^2$NICTA, Sydney, Australia ~~~ $^3$INRIA, Grenoble, France\\
%\{emmanuel.lochin,jerome.lacan, tuan.tran-thai@isae.fr\}@isae.fr, tournoux@gmail.com, Vincent.Roca@inria.fr
}
}

\maketitle

%\begin{abstract}
%  La partie 1) on présente les courbes de Tuan pour appuyer le fait qu'il est nécessaire d'utiliser des mécanismes d'effacements.  La section 2) pourrait reprendre une bonne partie de l'argumentation du Transaction on Multimedia avec un schéma Tetrys (si possible vu le format court). De toute façon c'est IETF draft donc une colonne en 3 pages faut être très succint.  La partie 3) quelques observations de mémoire concernant les expérimentations effectuées par Pierre-Ugo. Au sens, "There is no exsiting solution allowing to plug a CC with a EC. Our preliminary experiments have shown that TFRC is not directly compliant/efficient when EC is enabled because we observe that : blablabla" Pierre Ugo tu complètes, pas besoin de tartines.  How can we ensure that real-time communications are well-behaved with respect to other Internet applications while still providing good quality?  
%\end{abstract}

\section{Problem statement}
\label{sec:problemStatement}

\begin{figure*}[htb]
    \centering
        \includegraphics[width=0.5\textwidth]{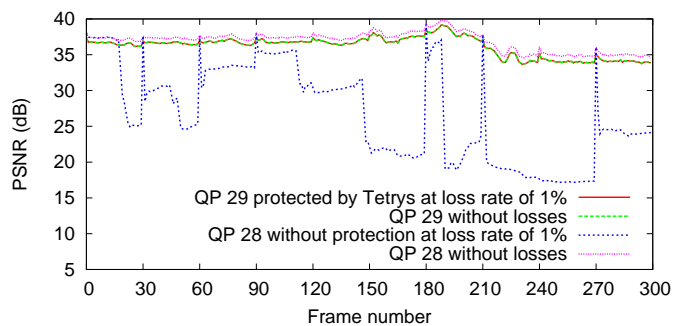}
 \caption{Video quality variation in presence of packet losses with and without erasure coding}
\label{fig:videoBehavior}
\end{figure*}

The use of real-time applications over the Internet is a challenging problem that the QoS epoch attempted to solve by proposing the DiffServ architecture. Today, the only existing service provided by the Internet is still best-effort. As a result, multimedia applications often perform on top of a transport layer that provides a variable sending rate. In an obvious manner, this variable sending rate is an issue for these applications with strong delay constraint. In a real-time context where retransmission can not be used to ensure reliability, video quality suffers from any packet losses. To illustrate this problem and motivate why we want to bring out a certain class of erasure coding scheme inside a multimedia congestion control protocol such as TFRC\cite{rfc5348}, we propose this simple simulation scenario: we encode a CIF 'Foreman' sequence of 300 frames with two values of Quantization Parameter (QP). For a Group of Picture (GOP) size of 30 frames, QP 28 and 29 result in H.264/AVC videos of $531.13$\,kb/s (Peak Signal to Noise Ratio (PSNR) of $36.9$\,dB) and 441.37 kb/s (PSNR of 36.2 dBs), respectively. A video with QP of 28 generates 16.9 $\%$ of redundancy bit rate compared to the one of 29. We drive the simulations with encoded video of QP 29 protected by an Erasure Coding (Tetrys mechanism \cite{TetrysJournal}) with redundancy of 12.5$\%$ and the one of 28 without protection. We randomly generate a loss rate of $1$\,\%. In Fig. \ref{fig:videoBehavior}, we observe a strong quality degradation of the video without protection (blue line compared to purple line) while Erasure Coding recovers all losses within one way End-to-End delay constraint of 150 ms (red and green lines are identical). This implies that a slightly higher quality without protection suffers from any losses while a slightly lower quality video with protection can sustain up to a certain loss rate (3\% in these simulations) depending on its redundancy. As a matter of fact, erasure-coding seems mandatory to enable a fair real-time video quality.

\section{How to cope with these losses?}

Currently, there are two kinds of reliability mechanisms based respectively on retransmission and redundancy schemes. Automatic Repeat reQuest (ARQ) schemes recover all lost packets thanks to retransmissions.
This implies that the recovery delay of a lost packet needs at least to wait one supplementary Round Trip Time (RTT) and this supplementary delay might exceed the threshold above a real-time application 
considers a packet outdated.

A well-known solution to prevent this additional delay is to add redundancy packets to the data flow. This can be done with the use of Application Level Forward Error Correction
(AL-FEC) codes. AL-FEC codes are FEC codes for the erasure channel where symbols (i.e. packets) are either received without any error or lost (i.e. erased) during transmission.
The addition of $n-k$ repair packets to a block of $k$ source packets allows to rebuild all of the $k$ source packets if a maximum of $n-k$ packets are lost among the $n$ packets sent.
In practice, only Maximum-Distance Separable codes (MDS), such as Reed-Solomon codes \cite{rfc5510}, have this optimal property, whereas other families of codes, such as LDPC \cite{rfc5170},
need to receive a few number of extra symbols in addition to the $k$ strict minimum.
However, if more than $(n-k)$ losses occur within a block, decoding becomes impossible.
In order to increase robustness (e.g. to tolerate longer bursts of losses), the sender can choose to increase the block size (i.e. the $n$ parameter) with the price of an increase of the decoding delay in case of erasure.
In order to improve robustness while keeping a fixed delay, the sender can also choose to add more redundancy while keeping the same block size with the price of a decrease of the goodput (which is not necessarily affordable by the application).
These trade-off between \textit{packet~decoding~delay; block~length} and \textit{throughput} are, for instance, addressed in \cite{flexibleRS09}.  
Another approach is proposed in \cite{martinian2004} where the authors use non-binary convolutional-based codes. They show that the decoding delay can be reduced with the use of a sliding window, instead of a block of source data packets, to generate the repair packets. However, both mechanisms do not integrate the receivers' feedbacks and thus, cannot provide any full reliability service.
Finally, an hybrid solution named Hybrid-ARQ which combines ARQ and AL-FEC schemes is often used.
This is an interesting solution to improve these various trade-off \cite{sahai:feedback08}.
However, when retransmission is needed, the application-to-application delay still depends on the RTT which might be not acceptable with real-time applications.

The novel coding scheme, named Tetrys \cite{TetrysJournal}, that we propose to introduce inside a congestion control protocol totally departs from the above schemes. 
Unlike current reliability methods, this on-the-fly coding scheme allows to fill the gap between systems without retransmission and fully reliable systems by means of retransmissions.
Tetrys needs only an average PLR to perform as its configuration does not depend of the size of the burst of losses.
The main properties of Tetrys is (1) to be tolerant to any burst of source, repair or acknowledgement losses, as long as the amount of redundancy exceeds the packet loss rate (PLR),
and (2) the lost packets are recovered within a delay that does not depend on the $RTT$, which is a key property for real-time applications.
We believe that these properties make Tetrys a good reliability candidate for a congestion control protocol such as TFRC \cite{rfc5348}.

\section{Issues for the Congestion Control}
\label{sec:congestionControl}

%Erasure codes brings a significant improvement when the protected applications have strong delay constraints. 
The use of erasure codes as a reliability mechanism requires to identify the key properties of the congestion control that they have to be paired with. 
The next two sections detail the relevant properties of the current congestion control and then describes the potential solution.
\subsection{Congestion control and real-time variable bit rate}
%Adapting the erasure code's configuration to maximize the tradeoff between delay, residual loss rate and the coding overhead is a challenging task. 
%Their usage to protect real-time traffic over the internet requires to take in account the congestion control which adapts the sending rate and ensure a fair share of the bandwidth with the other flows. Towards this end, 
TCP have been long know as an unsatisfying solution for carrying real-time traffic. Indeed, in case of a packet loss, the following packets will be stored in the receiver's windows until the loss is recovered, thus inducing a significant delay in the communication. The window-based congestion control of TCP itself has an impact on the application. Sending a packet requires to wait for an acknowledgement which, in case of losses or slow-start, tends to arrive by burst hence increasing the delay at the sender side. Finally, despite the fact that TCP's AIMD reaches a steady-state, its saw-tooth behavior prevents the application to adapt its bit rate. Furthermore, the buffering at the sender side might overtake the delay constraint of the application. As a result, TCP would support real-time traffic with a fair-share at least twice bigger than the source bit rate~\cite{streamingOverTCP}. For all these reasons, the support of real-time applications has turned towards protocols allowing out-of-order delivery and rate-based congestion control. 

TFRC \cite{rfc5348} is a congestion control mechanism designed to allow applications that use fixed packet size to compete fairly with TCP flows using the same packet size \textit{e.g.} 1500 bytes. If some real-time applications such as VoIP find a satisfying solution in TFRC, the applications that produce variable bit rate and variable packet size experiences severe performance issues when their sending rate is controlled by TFRC. This is for instance the case of video-conferencing where the stringent delay constraint and network capacity prevent the use of CBR codecs. 
%Due to the time constraints, such codecs encode frames using either only the information of the current frame (I-Frames) or only the difference with the previously encoded frames (P-Frames). To prevent the decoding errors to propagate through numerous frames, 
The video is encoded by independent group of picture (GoP) composed of one I-Frames followed by P-Frames with the size of resulting I-Frames being much larger the size of the P-Frames. 
%Depending on the encoding parameters, the size of I-Frames is likely to be several time bigger the size of P-Frames hence generating a burst of packets to transmits at the beginning of each GoP. 
As TFRC acts as a token bucket, the burst of packets induced by the I-frames has to be queued at the sender side before it can be entirely sent. The increased delay impairs on both the interactivity and the video quality in case of stringent delay constraint. The usual way to counter this drawback is to use padding and constantly transmit at the I-Frames rate. Obviously, it requires the fair-share to be much bigger than the bit rate of the of the application and it reduces the overall network goodput. Another drawback of TFRC is the packet size that is assumed to be fixed and similar to the packet size of the concurrent TCP flows. Smaller packets (such as the one generated by P-Frames) are counted as full size packet and the required fair-share is much bigger than the actual rate generated by the codecs.

Few work have been proposed enable VBR sources in TFRC. In~\cite{sathiaseelanF11}, the authors highlight the mismatch between the media rate and TFRC sending rate and propose to combine Faster Restart to TFRC to better support bursty applications. Although their study shows a certain improvement in terms of quality of experience for the user, the overall gain obtained is quite limited.
The issue induced by the packet size is also partially tackled in TFRC-VP~\cite{rfc4828} which proposes to send small packets while being fair with TCP flows sending MTU sized packets. 

\subsection{Coupling congestion control, real-time application and erasure coding}

To the best of our knowledge, there is no transport protocol allowing source burstiness and variable packet size and hence the improved compression and interactivity allowed by modern codecs. %As we have seen, erasure code and specifically on-the-fly erasure code allows a significant improvement in the quality of such applications but their performance varies with the available throughput and packet rate. %
While remaining TCP-friendly, the ideal solution should not to delay the sending of the data and the packet size should be arbitrary small. As a side effect, it would significantly ease the dimensioning of the erasure code in charge of loss recovery. We are aware that such a freedom in the packet timing might be hard to achieve -if not orthogonal to TCP-friendliness- and hence it should be bounded. Towards this end, the equation based congestion control TFMCC-VP~\cite{tfmcc-vp} allows to trade packet size for packet rate while remaining fair with TCP. Configuring erasure code is a multi-dimensional problem involving packet rate, decoding delay, redundancy ratio, loss rate and correction capability. The higher the packet rate is, the more the correction capability or the required redundancy ratio will be. With TFRC, when congestion occurs, the packet rate reduction requires more redundancy to achieve the goal in terms of delay and loss recovery. With TFMCC-VP \cite{tfmcc-vp}, instead of adding a new unknown parameters that will impair the correction capability of the erasure code, the packet rate can remain the same. As emphasized in~\cite{TetrysJournal}, on-the-fly erasure codes have a strong advantage compared to erasure block codes as for a given redundancy ratio, they shows the same performance than the best configuration among the erasure block codes. We consider that the robust configuration of on-the-fly codes combined with the fixed (or configurable) packet rate of TMFCC-VP allows a promising solution to enable congestion control for interactive real-time communication.

%The packet size and packet rate trade-off can be adapted to fit the application's packet size, the maximum packetisation delay and the packet rate required to by the optimal erasure code configuration. As emphasized in~\cite{TetrysJournal}, on-the-fly erasure codes are particularly adapted to VBR sources and impredictable and variable loss rate give them a significant advantage compared to block codes. Indeed, if the average loss rate and the delay constraint remain stable, the performance of on-the-fly code for a given redundancy budget will depends on the packet rate. If the available throughput decrease, the packet size can be reduced to keep the same packet rate and hence the same performances. The trade-off between packet rate, coding ratio and application throughput need a deep investigation but this proposal would allow a to combine realtime application, erasure codes and congestion control without requiring any change to the network infrastructure or sending more packets than the fair-share.

%We also consider that this is not the only design choice that can be lead to a satisfying solution. \textcolor{red}{DELETE ME IF YOU WANT : For instance, in the case of realtime video, the burstiness of the source can be mitigated by using only P-Frames. In order to limit the propagation of errors in the decoded video, it requires an erasure code that correct all the losses under strong delay constraints and a the video decoder should integrate the information brought by the packets that were missing then recovered.}

In the context of video streaming with loose delay constraints, the authors of~\cite{SeferogluTMM10} suggest that the redundancy should not be part of the congestion-controlled rate. Their erasure block code configuration scheme predicts the loss rate and sends the minimum coding rate sent in addition to TFRC's rate. However their scheme is not appropriate for strong delay constraint applications. The proposal of~\cite{Mahajan_eatall} is pushing further this idea and suggests that the redundancy packets should be marked and discarded following an AQM drop precedence policy. Despite the deployment of such scheme is questionable, we believe that solutions that send redundancy packets outside the congestion-controlled flow deserve to be investigated.

%\section{Check}
%\begin{itemize}
%\item Objective : erasure coding congestion controlled : Done first para
%\item Questions to solve (need advice and IAB position)
%    \begin{itemize}
%    	\item Redundancy outside or inside the congestion controlled flow : Done with the antepenultieme paragraph ~\cite{SeferogluTMM10}
%	\item Problem of some hypothesis of multimedia congestion control protocol : TFRC (Pierre-Ugo says and Pierre-Ugo writes ;)  Done for TCP and TFRC
%	\item Possible use of TCP-Friendly Multicast Congestion Control (TFMCC) instead of TFRC ? Done
%	\item Good point : TFRC feedback policy compliant with on-line coding scheme \textcolor{red}{NOT DONE}
%   \end{itemize}
%\end{itemize}

%- Idea : to bring out online erasure-coding inside a CC to improve the transport of real-time video

%\section{Conclusion}

\bibliographystyle{plain}
\bibliography{biblio}

\end{document}